# Fabrication of hexagonally ordered nanopores in anodic alumina: An alternative pretreatment

**K.M. Alam, A.P. Singh, S.C. Bodepudi, S. Pramanik**
Department of Electrical and Computer Engineering, University of Alberta, Edmonton, Canada

E-mail: pramanik@ece.ualberta.ca

**Abstract.** Anodic aluminum oxide (AAO) or anodic alumina template containing hexagonally ordered nanopores has been widely used over the last decade for the development of numerous functional nanostructures such as nanoscale sensors, computing networks and memories. The long range pore order requires the starting aluminum surface to be extremely smooth. Electropolishing is the most commonly used method for surface planarization prior to anodization. While prevalent, this method has several limitations in terms of throughput, polishing area and requirement of special experimental setups, which introduce additional speed bottlenecks in the intrinsically slow AAO-based nanofabrication process. In this work we report a new generation of the so-called "chemical polishing" approach which circumvents these stumbling blocks in the pretreatment phase and offers a viable, simpler, safer and faster alternative to electropolishing. These benefits are obtained without sacrificing the quality of the final AAO template. In this work we have (a) identified the optimum parameter regime for chemical polishing and (b) determined process conditions for which a novel parallel nanoridge configuration self-assembles and extends over a distance of several microns. Such patterns can be used as a mask for fabricating nanocrossbars, which are the main structural components in myriad nanoscale memories and crosspoint architectures.

# 1. Introduction

Over the last decade, nanoporous anodic aluminum oxide (AAO) template has emerged as a versatile and low-cost platform for fabricating a multitude of nanostructures which exhibit intriguing device properties [1–3]. AAO is characterized by a homogeneous morphology of parallel pores which grow perpendicular to the template surface with a narrow distribution of diameter, length and interpore spacing, all of which can be easily controlled by suitably choosing the anodization parameters such as pH of the electrolyte, anodization voltage and duration of anodization [1,2]. A wide range of materials can be deposited (or grown) selectively within the pores using various techniques such as electrodeposition [4,5], sputtering [6], thermal evaporation [7], pulsed laser deposition [8,9], chemical vapor deposition [10], sol–gel molding [11], filtration [12] and high temperature impregnation [13]. Such templates can also be used in conjunction with other top-down or self-assembly techniques to produce higher order nanofabrication protocols [14].

The steady state pore formation (and their self-assembly in a hexagonal pattern) results from a balance between two competing processes [1,3]: (a) electric field assisted dissolution of aluminum oxide (or alumina) at the oxide/electrolyte interface and (b) formation of alumina at the oxide/aluminum interface. The pore formation process can be made spatially periodic by patterning an extremely smooth aluminum substrate prior to anodization. Long range pore ordering is desirable for myriad applications such as nanoscale memories and photonic crystals. Lack of surface smoothness prevents self-organization of the pores and hence virtually all experiments that employ AAO start from a pretreatment step in which the aluminum substrate is first polished to attain a desired degree of surface smoothness. This smooth surface is then either subjected to (a)multiple anodization-and-etching steps [15] or (b) a pre-patterning stage followed by a single step anodization [16–19] to achieve long range pore order. The former method results in a quasi-monodomain structure whereas the latter produces a perfectly ordered nanopore lattice. In literature, electropolishing is the most commonly used surface planarization technique (see, for example, [15–29]). The so-called L1 solution (a mixture of 62 cm$^3$ perchloric acid, 700 cm$^3$ ethanol, 100 cm$^3$ butyl cellusolve and 137 cm$^3$ distilled water) has been studied extensively since it produces different nanoscale patterns under different electropolishing conditions [20,21]. Several other electrolytes such as perchloric acid–ethanol mixture [15,17,18,22,23,25,26,29] and a mixture of phosphoric acid, sulphuric acid, water (or chromic oxide) [24,27,28] have also been reported. Interestingly, perchloric acid is a common ingredient in most cases and produces better smoothness compared to the solutions that do not contain perchloric acid [20,27].

While popular, electropolishing (with a solution that contains perchloric acid) comes with several practical disadvantages all of which originate from the fact that hot perchloric acid is a hazardous chemical which requires special handling. For example, during electropolishing the electrolyte gets heated and hot perchloric acid is well known for its notoriously strong oxidizing properties and can also become unstable. This requires continuous cooling of the electrolyte during polishing. Even storing of these electrolytes require special "perchloric acid fume hood" with self-contained wash-down units to inhibit formation of metallic perchlorate crystals on the walls of the fume hood and the exhaust ductwork. These crystals are highly unstable and shock-sensitive and may ignite or detonate under certain conditions.

Electropolishing is mainly a serial process and with commercial electropolishers (such as Buehler Electromet-4) only one sample can be electropolished at a given time resulting in a slow throughput. Using home-built electropolishing setups it is possible to simultaneously electropolish multiple aluminum substrates in parallel in an electrolyte, but again this generates larger amount of heat (compared to the single substrate case) requiring highly efficient heat removal from the system. Ultimately the maximum number of substrates that can be electropolished simultaneously will be limited by the cooling capacity of the system. Third, the area of electropolishing is often limited in commercial electropolishers (e.g. the exposed diameter is ~2 cm for Buehler). Again, large area polishing can be achieved by home-made setups but at the cost of higher current and concomitant evolution of larger amount of heat. We note that one of the intrinsic drawbacks of AAO based nanofabrication is its slow processing speed. Recently few approaches have been proposed for accelerating this process [23,28]. However, the factors mentioned above indirectly introduce additional speed bottlenecks which thwart large scale production of high quality AAO templates. Thus an alternative fast polishing technique that will avoid these disadvantages and at the same time produce AAO templates of comparable quality is highly desirable.

In this work we draw renewed attention to the so-called "chemical polishing" process, which was originally invented in the 1930s [30]. This method has been rarely used in the development of nanostructures even though it allows parallel processing of large number of aluminum substrates with arbitrary areas and not necessarily requires perchloric acid. One early work [31] employed a variation of this technique for structural investigation of alumina films. In this report, the chemical polishing process is characterized by the use of a mixture of nitric and phosphoric acid (note the absence of perchloric acid) and an elevated solution temperature. However, use of this process in the AAO literature is rare (if any), mostly because of difficulties in controlling the surface smoothness and poor quality of the AAO template. We note that ref. [32] studied a similar chemical polishing approach, albeit with different polishing reagents and conditions. In this study, the chemical polishing produced inferior surface smoothness as compared to the electropolished surface. Ref. [29] reported yet another chemical polishing approach which produced disordered porous structures with adverse effects on the parallelism of the pores. In our work we demonstrate the opposite behavior i.e. we first show that the chemical polished surface is more smooth compared to the electropolished surface, if the polishing parameters lie in a certain window. The maximum size of the aluminum substrate is only limited by the size of the bath and we have successfully (chemical) polished substrates as large as 10 cm×10 cm. After chemical polishing, the entire surface is shiny and mirror-like and appears almost identical to the electropolished surface. Next we will show that after anodization of chemically polished surface, the parallelism of the pores is preserved and the template quality is comparable to that of electropolishing case. Thus the proposed method is a viable, simpler, safer, faster alternative for fabrication of good quality AAO templates. Surprisingly, under certain polishing conditions we observed self-assembly of highly ordered nanoridge patterns. Such morphologies can be used as mask to produce nanoscale interconnects in crossbar architecture.

This article is organized as follows. In Section 2 we describe the experimental procedure. The results and discussions will be presented in Section 3. We conclude in Section 4.

## 2. Experiment

We have used 2 cm×2 cm coupons of high purity unpolished aluminum (99.997%, Alfa Aesar) with thicknesses 250 μm and 100 μm. Both thicknesses have shown nominally identical results and in this article we will discuss the data obtained from 250 μm thick samples only. Larger area samples (10 cm×10 cm) were also studied and no significant difference has been observed. High purity aluminum is required to avoid potential breakdown phenomena and formation of cracks during the oxide formation (anodization) process. These unpolished samples have been subjected to the proposed nonstandard chemical polishing step. The etchant consists of 15 parts of 68% nitric acid and 85 parts of 85% phosphoric acid. This recipe completely avoids perchloric acid and the related hazards discussed before. The etching is performed for various durations (2–7 min for the temperature range 45 °C–95 °C). After etching, the samples are neutralized in 1 M sodium hydroxide for 20 min. Surface topography for each polishing condition has been studied extensively by atomic force microscopy (next section) to determine the optimum conditions.

Next, we prepared standard electropolished specimens to compare the quality of surface finish with the above-mentioned chemical polish method. There exist numerous recipes of electropolishing (e.g. [15–29]). In this work we choose the method of refs. [20,21], which has been thoroughly characterized in literature. In this method, the etchant consists of 700 $cm^3$ ethyl alcohol, 100 $cm^3$ butyl cellusolve, 62 $cm^3$ perchloric acid and 137 $cm^3$ distilled water. We employ a commercial Buehler Electromet-4 polisher/etcher apparatus with a polishing area ~2 cm in diameter. In this setup, the aluminum foil acts as the anode and remains in contact with the L1 electrolyte. Electropolishing is carried out under a dc bias of 45 volts for 10 s. Under these conditions, electropolishing current density is ~3 $A/cm^2$ which results in evolution of considerable amount of heat. The electrolyte is cooled down by passing cold water through the cooling coils which are in physical contact with the electrolyte. Cooling the electrolyte is necessary for two reasons: (a) using hot electrolyte for polishing results in visible imperfections and defects on the surface; and more importantly (b) hot perchloric acid is potentially explosive and can cause safety hazards. The electrolyte is vigorously stirred during electropolishing in order to prevent material congregation on the aluminum substrate. It is to be noted that the above-mentioned parameter set for electropolishing has been chosen from early work in this area which extensively characterized surface smoothness as a function of various electropolishing parameters. In these studies, longer electropolishing time did not show significant improvement in surface roughness [21]. However, in principle it may be possible to obtain better surface smoothness for a separate set of electropolishing conditions but this is beyond the scope of present work. Even in such case, the purely chemical method reported here will still remain a simpler, safer and faster alternative.

Finally, we perform multistep anodization on the chemical polished samples to examine the quality of the porous structures. Multistep anodization is a standard process for fabrication of ordered nanopores on the alumina template and has been investigated by many authors (e.g. [15]). Briefly, a polished piece of aluminum is anodized for short time (~15 min, 3% oxalic acid and 40 V dc at room temperature) which results in a thin layer of nanoporous oxide on the metal. These pores are distributed somewhat randomly on the surface. This thin oxide is dissolved out in hot (60 °C) chromic– phosphoric acid. Then, the sample is re-anodized (under the same conditions) for a long time (10–12 h) to form a thick layer of nanoporous film on the surface. Long time anodization reduces the defects and dislocations in the sample. This oxide layer is again removed by dissolving in hot chromic–phosphoric acid. Removal of this oxide leaves an array of highly ordered "dimples" on the metal surface. For subsequent (and final) anodization process these dimples act as initiation sites for the pores. The final step of anodization is carried out for the appropriate duration (under same conditions as in the previous steps) to attain the desired film thickness. All anodization processes in this work have been carried out with a platinum mesh counterelectrode.

The anodized area is typically composed of several μm² size "domains". The defects and imperfections tend to accumulate at the domain boundaries. Near perfect pore ordering is obtained within each domain. The arrangement of pores varies slightly from one domain to another. Further improvement in pore ordering can be obtained via imprinting techniques, which are beyond the scope of this work. We show that after multistep anodization, it is possible to grow AAO templates with large area domains on chemical polished samples. The domain areas of these templates are comparable to those grown on electropolished substrates. Thus using chemical polishing we are not sacrificing the template quality (as compared to electropolishing) but are able to avoid the speed bottlenecks associated with electropolishing protocol. Thus chemical polishing approach is more appealing for fast fabrication of high quality AAO samples in a simple and less hazardous fashion.

The surface characterization has been performed using an Atomic Force Microscope (Asylum Research, MFP-3D) under ambient conditions using a standard tetrahedral silicon tip (Olympus, OMCLAC160TS- W2) located at the end of a silicon cantilever. The typical values of scan rate, force constant and resonant frequency were 1 Hz, 42 N/m, 300 kHz respectively. The radius of curvature of the tip is <10 nm including the aluminum reflex coating. Characterization of the nanopores after multistep anodization, has been performed by a JEOL 6301 F field emission scanning electron microscope.

## 3. Results and discussion

To characterize and compare the surface roughnesses under different polishing conditions, we use two statistical parameters [33–36]: (a) standard deviation ($\sigma$) of the distribution of surface heights and (b) height correlation length (l). The standard deviation is calculated using the formula [34,37]:

$$\sigma = \langle \sigma_0 \rangle = \left\langle \left( \frac{1}{N^2-1} \sum_{i=1}^{N^2} (z_i - \bar{z})^2 \right)^{\frac{1}{2}} \right\rangle \tag{1}$$

where $N^2$ is the total number of pixels in the image, $z_i$ is the height of the $i$th pixel, $\bar{z}$ is the average height of the image and $\langle \rangle$ indicates averaging over five non-overlapping areas of same scan size. However, $\sigma$ or the shape of the height distribution do not provide the information regarding spatial distribution of the heights. This can be obtained from the two-dimensional height autocorrelation function which has the following (unnormalized) form:

$$C(p,q) = \sum_{i,j} z(i,j) z(i+p, j+q) \tag{2}$$

where ($p$, $q$) indicates the so-called lag vector $\vec{\beta}$ and (i, j) is the position vector in the *xy* plane. The correlation length *l* is defined as [35]

$$C(l/2) = \frac{1}{e} C(0,0) \tag{3}$$

Physically, the correlation length indicates the distance (or lag$|\vec{\beta}|$) over which the "memory" of the initial height is retained. Beyond this distance, the two points on the surface can be considered as statistically independent. Computation of the normalized autocorrelation function ($C_0$) is performed by the following routine [38]:

$$C_0(p,q) = \frac{\sum_{i,j}\left[z(i,j)-\bar{z}\right]\left[z(i+p,j+q)-\bar{z}_{p,q}\right]}{\left\{\sum_{i,j}\left[z(i,j)-\bar{z}\right]^2 \sum_{i,j}\left[z(i+p,j+q)-\bar{z}_{p,q}\right]^2\right\}^{\frac{1}{2}}} \tag{4}$$

where $z(i, j)$ represents the image height at (i, j), $\bar{z}$ is the image height as defined before, and $\bar{z}_{p,q}$ is the mean height of the shifted image $z(i+p, j+q)$. The correlation length is defined as in Eq. (3). Using these two parameters ($\sigma$, $l$), we determine the optimum conditions for chemical polishing and compare this result with the electropolished case. The AFM images in this paper are the raw data, without any artificial filtering or smoothing.

Fig. 1a shows the AFM image of a typical as-received unpolished aluminum foil. The parallel trenches are readily observed, which result from the rolling of the aluminum foil. The height variation over the entire surface is ~378 nm and the standard deviation $\sigma_0$ is 47.5 nm. The depth profile along a direction transverse to the trenches is shown in Fig. 1b. Such roughness makes these unpolished substrates unsuitable for attaining long range order. This calls for a surface polishing step.

*3.1 Chemical polishing*

First we employ the chemical polishing procedure as described in Section 2. In contrast with the electropolishing case where polishing voltage (along with temperature and duration) plays a crucial role in determining surface features [20,21], here we have only two parameters (etchant temperature and duration of etching) to optimize. Fig. 2 shows the variation of $\sigma$ and $\langle l \rangle$ with temperature. Etching time has been kept fixed for five minutes since for longer etching, the whole aluminum substrate is dissolved away at high temperatures. We observe that $\sigma$ shows a pronounced minimum and $\langle l \rangle$ attains a maximum at 85 °C, indicating that the best surface smoothness is obtained at 85 °C. Similarly, we plot $\sigma$ and $\langle l \rangle$ as a function of etching time (for a fixed temperature of 85 °C) in Fig. 3. The optimum etching time is five minutes. Longer etching tends to dissolve away the entire substrate whereas shorter etching leaves rough features on the surface. The $\sigma$ values are typically bigger for larger scan sizes since larger areas contain more variations in height. The $l$ (and $\langle l \rangle$) values are, however, relatively independent of scan size, as expected.

Fig. 4 shows the surface morphology of a chemical polished aluminum foil, polishing being performed for 5 min at 85 °C. The parallel trenches of the unpolished specimen (Fig. 1) are no longer visible and the peak to peak roughness (over the entire area) has reduced to ~66.5 nm and $\sigma_0$=4.93 nm. If we take $\sigma_0$ as a guide, this data indicates almost an order of magnitude improvement in surface smoothness over the unpolished case. Note that both images have the same scale for surface heights.

*3.2. Comparison with electropolishing*

Next, we compare this method with the standard electropolishing results. Fig. 5 shows the surface morphology of an electropolished aluminum foil. Electropolishing has been performed according to the process outlined in Section 2. The peak-to-peak roughness is ~83.74 nm and $\sigma_0$=5.68 nm over an area of 20 μm×20 μm, which are significantly lower compared to the unpolished case but slightly higher than the chemical polished sample (Fig. 4). To further confirm this dependence of surface roughness on polishing conditions, we compare the height histograms of a different set of samples which exhibit (a) $\sigma_0$=68.4 nm (unpolished), (b) $\sigma_0$=6.9 nm (electropolished) and (c) $\sigma_0$=5.47 nm (chemical polished) for 20 μm scan size in Fig. 6. The chemical polished sample clearly shows the smallest standard deviation and indicates most smooth surface, which is consistent with the previous set of samples which correspond to Figs. 1, 4 and 5. The difference in surface roughness can also be seen immediately from Fig. 7 which compares the three dimensional presentation of the AFM topographic images shown in Figs. 1, 4 and 5. Polished specimens are observably smoother than the unpolished one. More importantly, the chemical polishing method reported above produces surfaces which are comparable or slightly better than the electropolished surfaces.

In Fig. 8 we compare the correlation lengths for these three surfaces. The unpolished aluminum shows a correlation length of 12 nm whereas electropolished and chemical polished samples have correlation lengths of 39 nm and 40 nm respectively. These data is consistent with the observation made above regarding the surface quality of the chemical polished specimens.

Table 1 summarizes the average standard deviation ($\sigma = \langle \sigma_0 \rangle$) and average correlation length ($\langle l \rangle$) for unpolished, electropolished and chemical polished specimens. The average has been calculated over five different areas with same scan size of 20 μm. For the electropolished surface, $\sigma$ =8.92 nm ($\langle l \rangle$=56 nm) whereas $\sigma$ ($\langle l \rangle$) for chemical polished samples is 7.1(62) nm. The $\sigma(\langle l \rangle)$ for unpolished sample has a much higher (lower) value of 70(12) nm. Thus we conclude that compared to electropolishing, chemical polishing can produce similar (or slightly better) surface smoothness under certain conditions. It now remains to be seen if chemical polishing has any adverse effect on pore formation and their regimentation.

*3.3. Multistep anodization on chemical polished samples*

Fig. 9 a (b) shows the top (cross-sectional) view of an AAO template fabricated by multistep anodization (Section 2) on a chemical polished aluminum substrate. The pores are clearly visible and form a well-regimented array within domains separated by defects and imperfections. To quantify the area of the perfectly ordered domain we follow the method used in [39]. Fig. 10 a and b show the schematic description and FESEM image of perfect hexagonally ordered pores respectively. One pore is chosen randomly as a reference and diagonal lines are drawn along three close packing directions. Within an ordered domain the pores will lie on these diagonals. If the diagonals have n pores, the area of the regular hexagon will be given by

$$S = \frac{3\sqrt{3}}{8} D_0^2 (n-1)^2, n = 3, 5, 7, \ldots$$

where $S$ is the area of the regular hexagon and $D_0$ is the interpore separation (~110 nm in our samples). We use this formula to calculate the most probable domain size in our samples. For this calculation we first determine the most probable value of $n$, following the steps described below [39].

First, we selected random pores on the 20 K magnification SEM images taken at different locations of the same sample. Next step was to draw three diagonal lines along three close packing directions. The diagonals are terminated at domain boundaries beyond which the pores no longer necessarily lie on the diagonal. We have repeated this process for 40 random pores in different locations on the sample. For each position there are three different $n$ values. Finally we plotted the histogram of all these $n$ values (total 40×3=120) in Fig. 10 c and performed the Gaussian fitting. The most probable value for $n$ is 21.16, as determined from this distribution. The average domain area, using the formula mentioned above, turns out to be 3.19 μm$^2$ with a standard deviation of 0.064 μm$^2$.

These numbers are comparable with previous studies (e.g. [39]) which characterized domain sizes after multistep anodization on electropolished specimens. This study reported a maximum domain area of 2.6 ± 0.11 μm$^2$ after rigorous preprocessing that includes extensive annealing combined with both chemical and electropolishing. Here we show that a comparable (or even better) porous structure can be obtained just by chemical polishing. This method is also faster and safer as described before. We also note that Fig. 9 is in sharp contrast with previous studies (e.g. [29]), which reported disordered pore arrangement as a result of chemical polishing. The current-time transients during anodization (not shown) are nominally identical to the electropolished case and the pore initiation process starts within the first few seconds. Thus the proposed chemical polishing method is a viable alternative for the standard electropolishing protocol. The domain size can be further improved by tuning the annealing parameters (temperature, time [39]) prior to the polishing step. Long range pore order can be achieved by using prepatterning techniques [16–19] on chemical polished surface.

Highly ordered nanopores of different diameters were successfully synthesized on chemical polished surfaces. For example, anodization using 0.3 M sulphuric acid and 25 V dc results in nanopore arrays with nominal pore diameter of 20 nm (Fig. 11). Similarly, anodization using 1.6 M malonic acid and 120 V dc produces nanopore arrays with nominal pore diameter of 200 nm. It is to be noted that chemical polishing is much more convenient than electropolishing especially when lateral anodization of aluminum substrate is required. This type of anodization has recently attracted significant attention since it opens the possibility of synthesizing three terminal devices using AAO technique [40,41].

*3.4. Evolution of nanoridges*

Finally, we would like to make an intriguing observation about the surface topography beyond the optimum polishing temperature. While the smoothest surface is obtained for an etchant temperature of 85 °C, nanoscale ridges begin to evolve at ~90 °C. Such a configuration is shown in Fig. 12. The peak to trough ratio is ~6 with a periodicity of ~0.5 μm. The ridges are parallel, extend over several microns, have a full width half maximum of ~0.1 μm. Such patterns can be potentially exploited as a mask to develop crossbars for nanoscale memory circuits [20,21]. The underlying mechanism and optimization of such ridge formation is currently under investigation and beyond the scope of this article.

## 4. Conclusion

In this work we have demonstrated a chemical polishing technique which offers a simpler, faster and safer route for aluminum pretreatment prior to anodization. This work is intended to draw renewed attention on the oft-overlooked chemical polishing method which has several advantages over the commonly employed electropolishing approach. We have performed extensive study of the surface roughness which shows that this method produces similar (or better) results compared to electropolishing. Additionally, this process (a) is parallel with fast throughput (i.e. multiple aluminum substrates with arbitrary areas can be processed simultaneously), (b) is amenable to large area polishing and (c) avoids perchloric acid and related hazards altogether. Ordered nanopores are formed after multistep anodization of chemical polished surfaces. Nanopore formation process is essentially identical to that in the electropolished case. Under certain polishing conditions parallel nanoridges are formed which can be further processed to fabricate parallel arrays of nanoscale crossbars. Such structures are the linchpins of many nanoscale memories and computing architectures.

**Acknowledgements**

This work has been supported by the Disruptive Technology Challenge program (TRLabs, Canada) and the Discovery Grant program (NSERC, Canada).

Table 1. $\sigma$ and $\langle l \rangle$ of the images for different polishing conditions. Chemical polishing has been done under the optimum conditions of 85 °C and 5 min (ref. Figs. 2 and 3). Scan size is 20 μm×20 μm.

| Roughness parameters(nm) | Unpolished | Electropolished | Chemical polished |
| --- | --- | --- | --- |
| $\sigma$ | 70 | 8.92 | 7.1 |
| $\langle l \rangle$ | 12 | 56 | 62 |

# References

[1] G. Sulka, Highly ordered anodic porous alumina formed by self-organized anodizing, in: A. Eftekhari (Ed.), Chapter 1 in Nanostructured Materials in Electrochemistry, Wiley-VCH, 2008.

[2] B. Kanchibotla, S. Pramanik, S. Bandyopadhyay, Self-assembly of nanostructures using nanoporous alumina templates, in: S. Lyshevski (Ed.), Chapter 9 in Nano and Molecular Electronics handbook, Taylor & Francis, 2007.

[3] S. Pramanik, B. Kanchibotla, S. Sarkar, G. Tepper, S. Bandyopadhyay, (to appear), Electrochemical Self-assembly of Nanostructures: Fabrication and Device Applications, in Encyclopedia of Nanoscience and Nanotechnology, Ed. H. Nalwa (American Scientific Publishers).

[4] T. Kline, M. Tian, J. Wang, A. Sen, M. Chan, T. Mallouk, Inorganic Chemistry 45 (2006) 7555.

[5] L. Liu, W. Zhou, S. Xie, L. Song, S. Luo, D. Liu, J. Shen, Z. Zhang, Y. Xiang, W. Ma, Y. Ren, C. Wang, G. Wang, Journal of Physical Chemistry C 112 (2008) 2256.

[6] W. Lee, M. Alexe, K. Nielsch, U. Gösele, Chemistry of Materials 17 (2005) 3325.

[7] D. Losic, J. Shapter, J. Mitchell, N. Voelcker, Nanotechnology 16 (2005) 2275.

[8] M. Cahay, K. Garre, J. Fraser, D. Lockwood, V. Semet, V. Binh, S. Bandyopadhyay, S. Pramanik, B. Kanchibotla, S. Fairchild, L. Grazulis, Journal of Vacuum Science and Technology B 25 (2007) 594.

[9] X. Gao, L. Liu, B. Birajdar, M. Ziese, W. Lee, M. Alexe, D. Hesse, Advanced Functional Materials 19 (2009) 3450.

[10] Z. Yuan, H. Huang, H. Dang, J. Cao, B. Hu, S. Fan, Applied Physics Letters 78 (2001) 3127.

[11] C. Goh, K. Coakley, M. McGehee, Nano Letters 5 (2005) 1545.

[12] M. Lahav, T. Sehayek, A. Vaskevich, I. Rubinstein, Angewandte Chemie. International Edition 42 (2003) 5576.

[13] L. Liu, S. Lee, J. Li, M. Alexe, G. Rao, W. Zhou, J. Lee, W. Lee, U. Gösele, Nanotechnology 19 (2008) 495706.

[14] H. Chik, J. Liang, S. Cloutier, N. Kouklin, J. Xu, Applied Physics Letters 84 (2004) 3376.

[15] H. Masuda, M. Satoh, Japanese Journal of Applied Physics 35 (1996) L126.

[16] H. Masuda, H. Yamada, M. Satoh, H. Asoh, M. Nakao, T. Tamamura, Applied Physics Letters 71 (1997) 2770.

[17] C. Liu, A. Datta, Y. Wang, Applied Physics Letters 78 (2001) 120.

[18] I. Mikulskas, S. Juodkazis, R. Tomasiunas, J. Dumas, Advanced Materials 13 (2001) 1574.

[19] K. Yasui, K. Nishio, H. Nunokawa, H. Masuda, Journal of Vacuum Science and Technology B 23 (2005) L9.

[20] S. Bandyopadhyay, A. Miller, H. Chang, G. Banerjee, V. Yuzhakov, D.-F. Yue, R. Ricker, S. Jones, J. Eastman, E. Baugher, M. Chandrasekhar, Nanotechnology 7 (1996) 360.

[21] R. Ricker, A. Miller, D.-F. Yue, G. Banerjee, S. Bandyopadhyay, Journal of Electronic Materials 25 (1996) 1585.

[22] L. Ba, W. Li, Journal of Physics. D. Applied Physics 33 (2000) 2527.

[23] W. Lee, R. Ji, U. Gösele, K. Nielsch, Nature Materials 5 (2006) 741.

[24] M. Tian, S. Xu, J. Wang, N. Kumar, E. Wertz, Q. Li, P. Campbell, M. Chan, T. Mallouk, Nano Letters 5 (2005) 697.

[25] N. Tasaltin, S. Ozturk, H. Yuzer, Z. Ozturk, Journal of Optoelectronic and Biomedical Materials 1 (2009) 79.

[26] A. Vazquez, R. Carrera, E. Arce, N. Castillo, S. Castillo, M. Moran-Pineda, Journal of Alloys and Compounds 483 (2009) 418.

[27] O. Jessensky, F. Müller, U. Gösele, Journal of the Electrochemical Society 145 (1998) 3735.

[28] K. Biswas, H. Matbouly, V. Rawat, J. Schroeder, T. Sands, Applied Physics Letters 95 (2009) 073108.

[29] P. Bocchetta, C. Sunseri, G. Chiavarotti, F. Di Quarto, Electrochimica Acta 48 (2003) 3175.

[30] Anodic Oxidation of Aluminium and its Alloys published by the, Aluminium Development Association, London, 1948.

[31] R. Alwitt, C. Dyer, B. Noble, Journal of the Electrochemical Society 129 (1982) 711.

[32] S. Ono, M. Saito, H. Asoh, Electrochimica Acta 51 (2005) 827.

[33] G. Simpson, D. Sedin, K. Rowlen, Langmuir 15 (1999) 1429.

[34] N. Almqvist, Surface Science 355 (1996) 221.

[35] C. Lui, L. Liu, K. Mak, G. Flynn, T. Heinz, Nature 462 (2009) 339.

[36] B. Bhushan, M. Nosonovsky, Scale effect in mechanical properties and tribology, in: B. Bhushan (Ed.), Chapter 16 in Nanotribology and Nanomechanics — An Introduction, Springer, 2005.

[37] M. Rao, B. Mathur, K. Chopra, Applied Physics Letters 65 (1994) 124.

[38] MATLAB normxcorr2 function.

[39] Z. Lu, Z. Xiao, P. CaoFeng, Z. Jing, Science in China. Series E: Technological Sciences 51 (2008) 1838.

[40] M. Gowtham, L. Eude, C.S. Cojocaru, B. Marquardt, H.J. Jeong, P. Legagneux, K.K. Song, D. Pribat, Nanotechnology 19 (2008) 035303.

[41] T.L. Wade, X. Hoffer, A. Mohammed, J.-F. Dayen, D. Pribat, J.-E. Wegrowe, Nanotechnology 18 (2007) 125201.

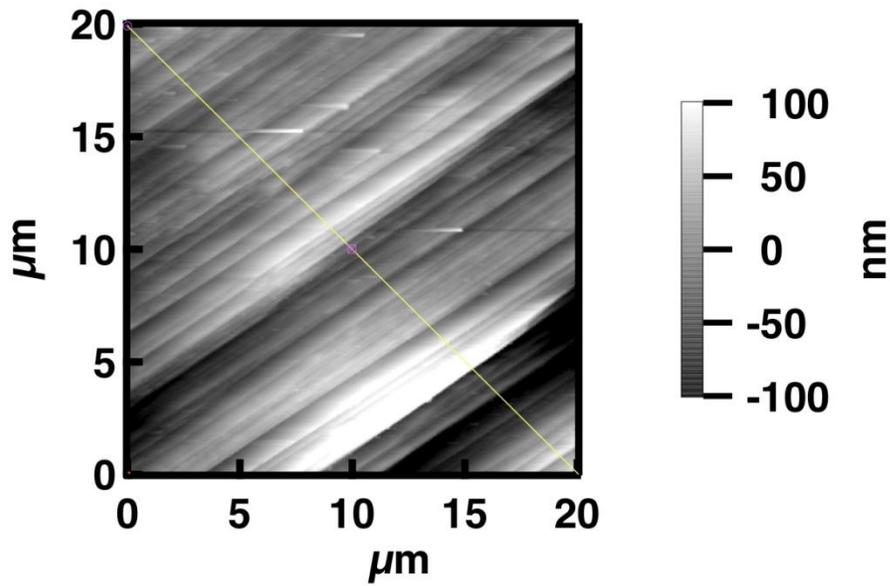

(a)

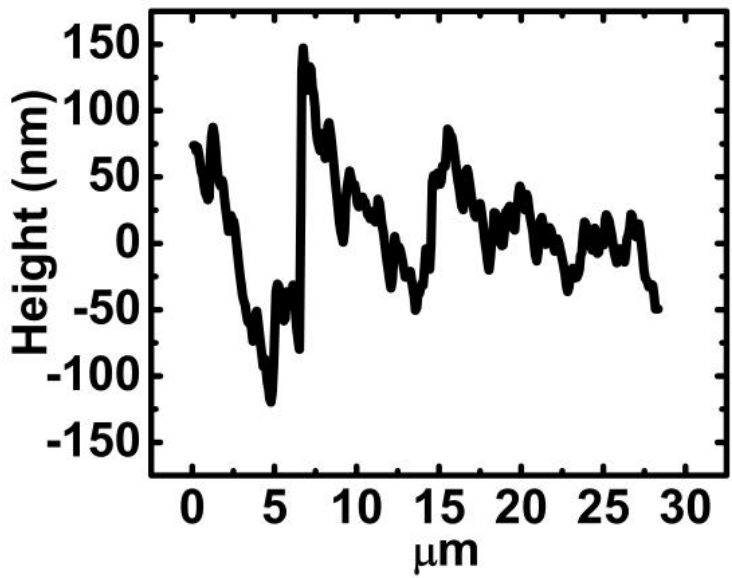

(b)

Fig. 1. (a) Two dimensional AFM topographic image of an unpolished sample for 20 μm scan size. The maximum height variation and standard deviation ($\sigma_0$) over the entire surface are 378 nm and 47.5 nm respectively. (b) Roughness profile along the diagonal line shown in (a). Along this line the peak-to-peak height variation is ~275 nm. Such roughness makes these samples unsuitable for attaining long range pore order after anodization

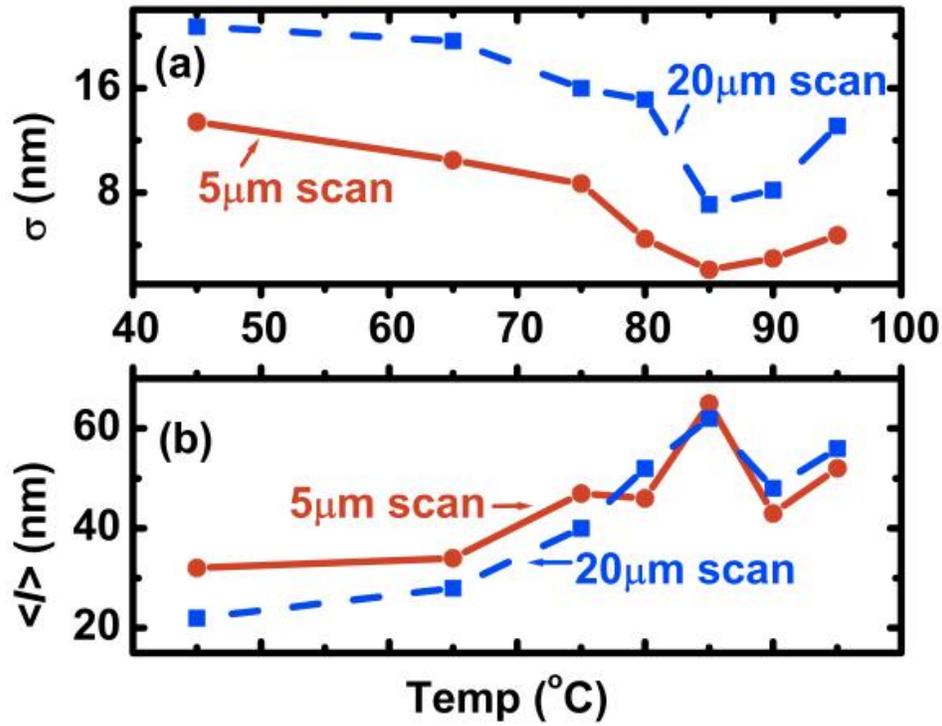

Fig. 2. Dependence of (a) standard deviation ($\sigma$) of the height distributions and (b) correlation length ($\langle l \rangle$) on temperature. Each data point is the average of five non-overlapping areas of same scan size. Etching time is five minutes in each case. The lines through the data points are guides to the eye.

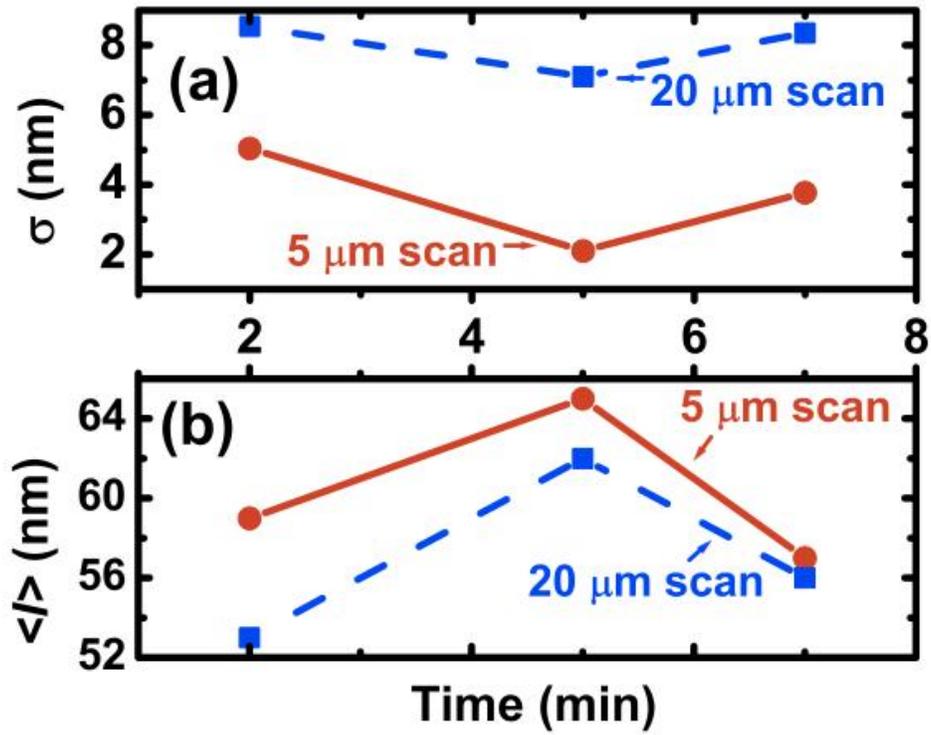

Fig. 3. Dependence of (a) standard deviation of the height distributions (σ) and (b) correlation length (⟨ l ⟩) on etching time. Each data point is the average of five non-overlapping areas of same scan size. Etching temperature is 85 °C in each case. The lines through the data points are guides to the eye.

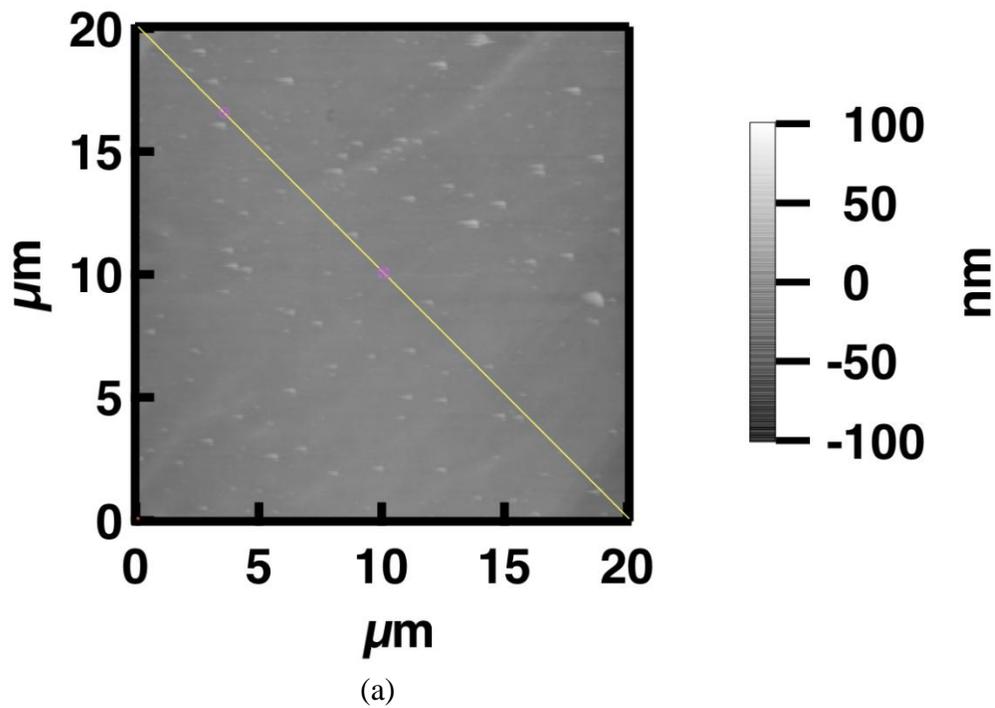

(a)

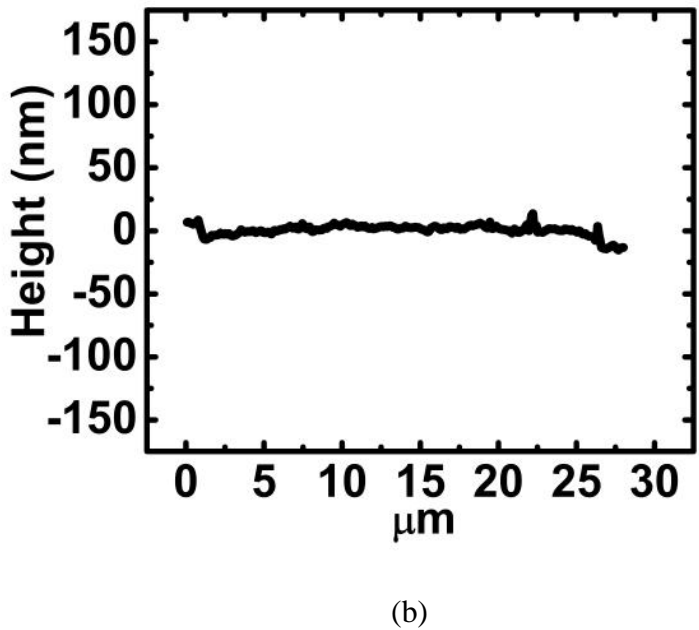

(b)

Fig. 4. (a) Surface morphology of a chemical polished sample for polishing temperature of 85 °C and duration of 5 min. The peak-to-peak height variation and the standard deviation ($\sigma_0$) over the entire surface (20 μm×20 μm) are 66.5 nm and 4.93 nm respectively. (b) Height trace along the diagonal line in (a). The peak-to-peak height variation along this line is ∼28.52 nm. The height scales are identical to that in Fig. 1.

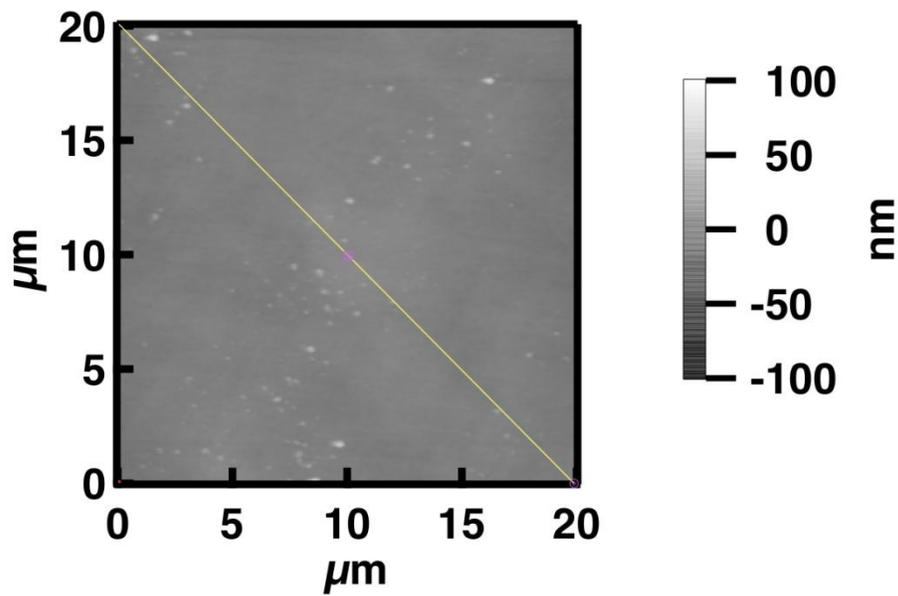

(a)

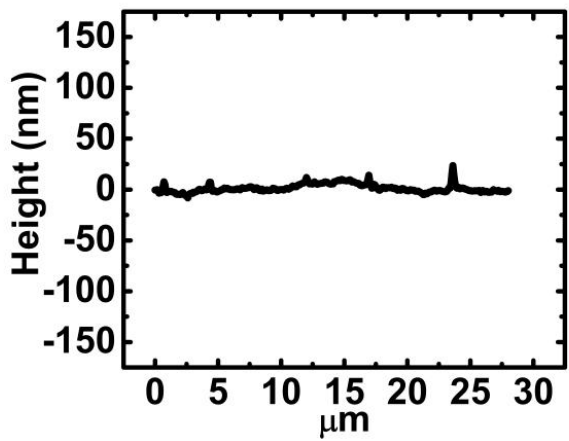

(b)

Fig. 5. (a) Surface morphology of an electropolished sample. The polishing parameters are described in Section 2. The peak-to-peak variation and standard deviation ($\sigma_0$) over the entire surface (20 μm×20 μm) are 83.74 nm and 5.68 nm respectively. (b) Height trace along the diagonal line in (a). The peak-to-peak height variation along this line is ~31.83 nm. The height scales are identical to those in Figs. 1 and 4.

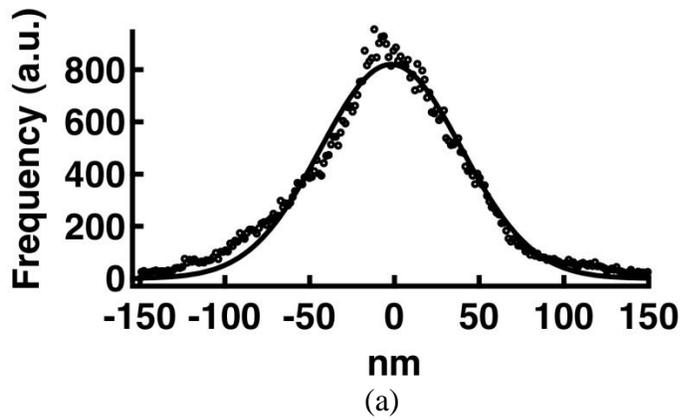

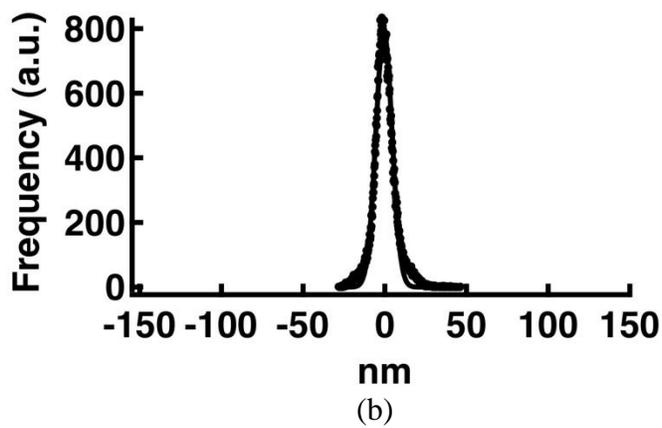

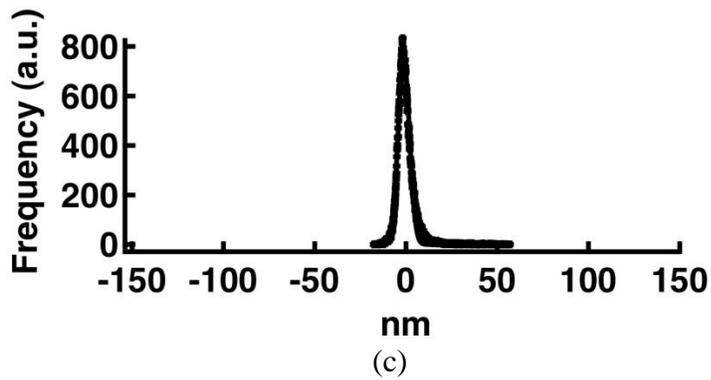

Fig. 6. Height histograms for (a) unpolished, (b) electropolished and (c) chemical polished specimens for a scan size of 20 μm. The continuous curves represent Gaussian fits with standard deviations ($\sigma_0$) of 68.4 nm, 6.9 nm and 5.47 nm respectively. The polishing conditions are the same as in previous figures.

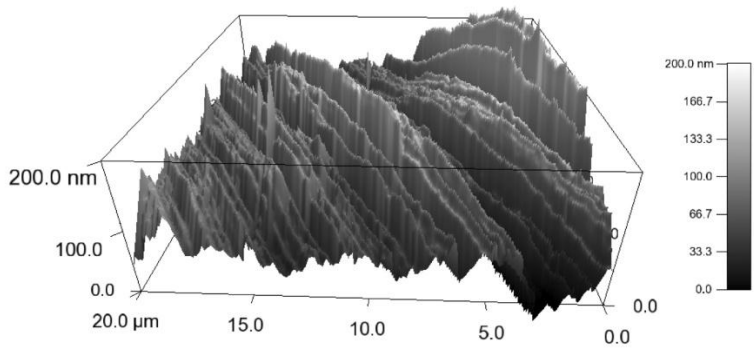

(a)

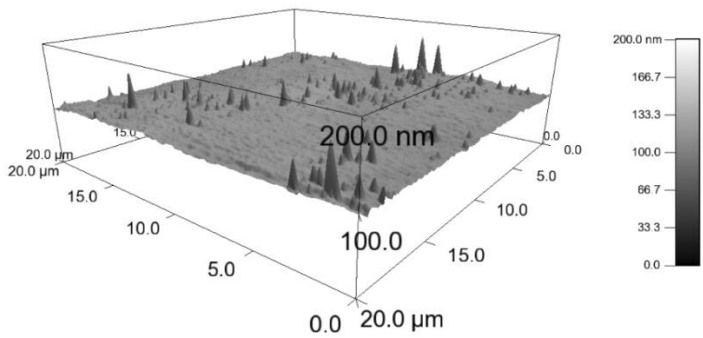

(b)

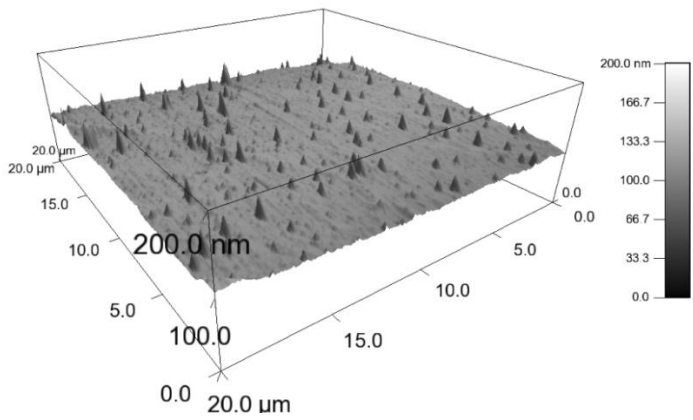

(c)

Fig. 7. Three-dimensional representations of the AFM topographic data for (a) unpolished (Fig.1), (b) electropolished (Fig. 5) and (c) chemical polished (Fig. 4) samples.

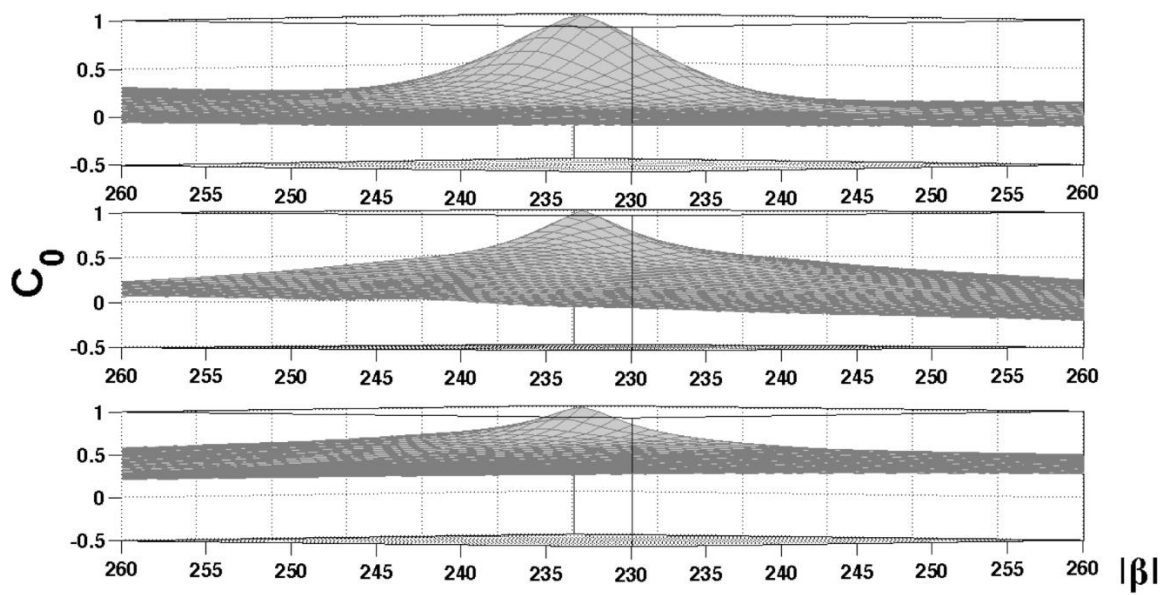

Fig. 8. Normalized height autocorrelation functions for (top) unpolished (Fig. 1), (middle) electropolished (Fig. 5) and (bottom) chemical polished (Fig. 4) samples. The correlation lengths (measured from $C_0=1$) are 12 nm, 39 nm and 40 nm respectively. The lag vector ($\vec{\beta}$) in the xy plane is in nm. The polishing conditions are the same as in previous figures.

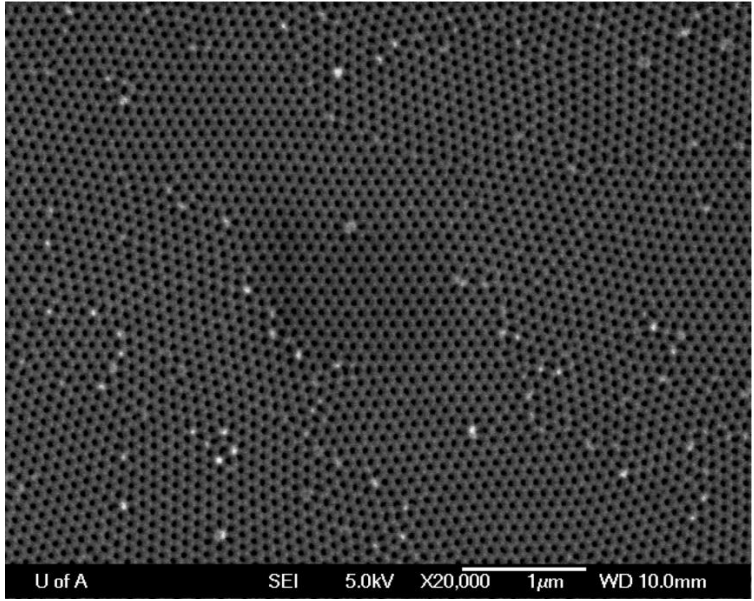

(a)

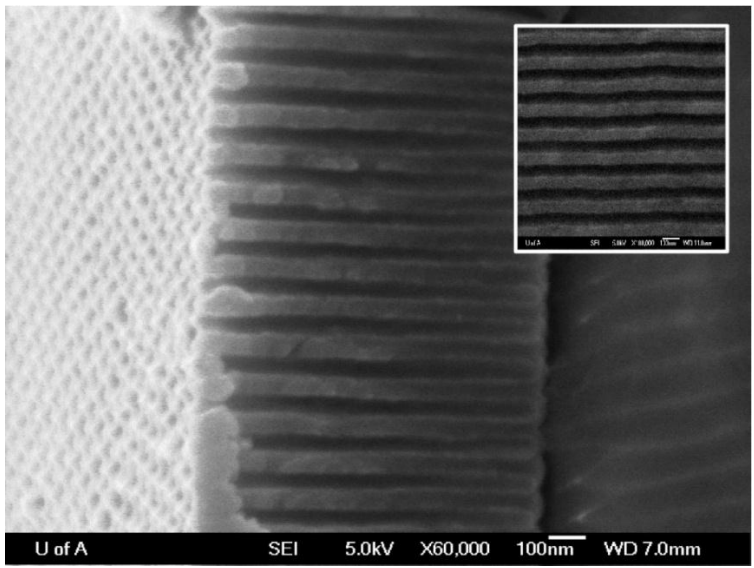

(b)

Fig. 9. (a) FESEM image of the top surface after multistep anodization on a chemical polished (85 °C, 5 min) sample. Before chemical polishing, the aluminum foil was annealed at 500 °C for three hours. (b) A FESEM image of the cross section of the template. Pore ordering and pore straightness are simultaneously visible. The inset shows a magnified image of the cross-section.

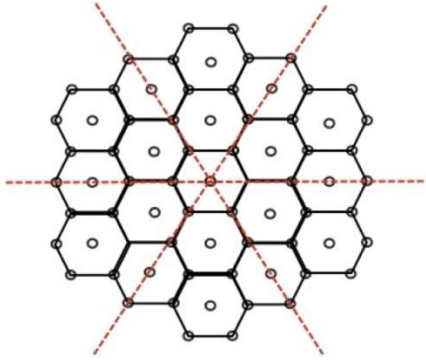

(a)

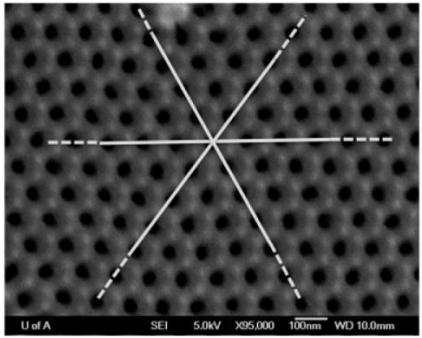

(b)

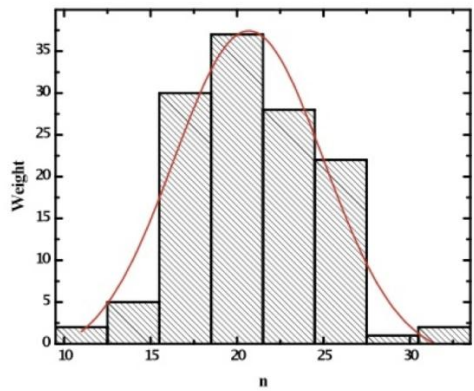

(c)

Fig. 10. (a) Schematic representation of the hexagonal pore arrangement. A random pore at the center has been chosen as the reference and three diagonal lines along the close packing directions are shown. (b) A FESEM image of a part of a hexagonally ordered domain. (c) Distribution for n (the number of pores lying on a diagonal line).

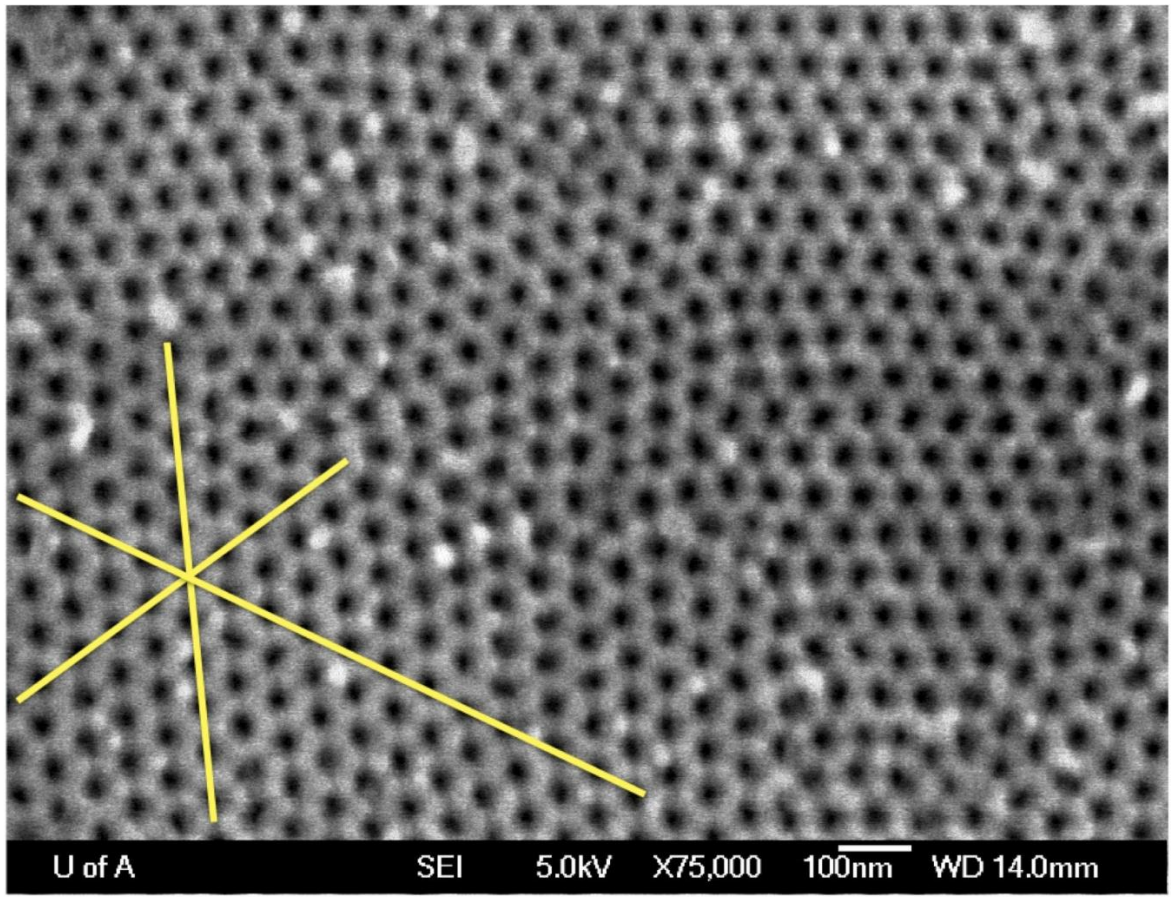

Fig. 11. A FESEM image of nanopore arrays with nominal pore diameter of ~20 nm. The diagonal lines indicate the hexagonal close packing directions.

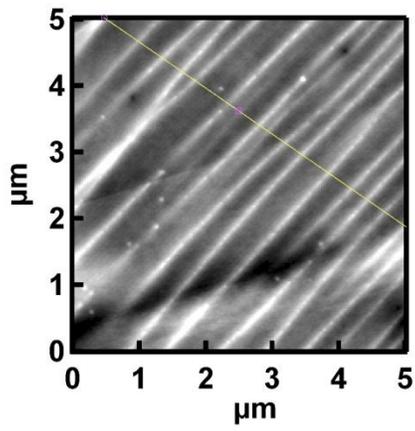

(a)

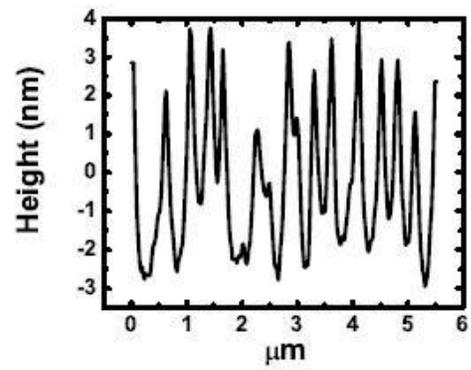

(b)

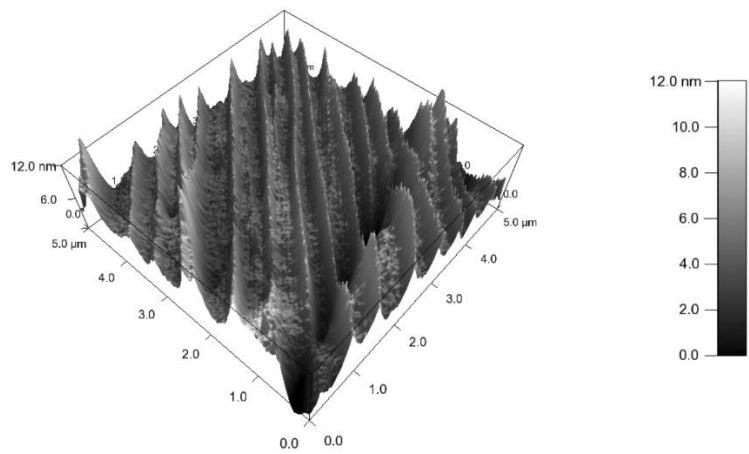

(c)

Fig. 12. (a) AFM image, (b) line plot and (c) three dimensional topography of a chemical polished sample at a polishing temperature of 90 °C. The thickness (FWHM) of the ridges are ~100 nm. The periodicity is ~0.5 μm and the peak to valley ratio is ~6.